\documentstyle[epsfig]{aipproc}
\begin{document}
\title{Laser Shaping and Optimization of the Laser-Plasma Interaction}
\author{Anatoly Spitkovsky$^*$ and Pisin Chen$^{\dagger}$}

\address{ 
$^*$ Department of Physics, University of California at Berkeley, Berkeley, CA 94720 \\
$^{\dagger}$ Stanford Linear Accelerator Center, Stanford University, Stanford, CA 94305}
\maketitle
\begin{abstract}
The physics of energy transfer between the laser and the plasma
in laser wakefield accelerators is studied. We find that wake excitation
by arbitrary laser shapes can be parameterized using the total pulse energy and
pulse depletion length. 
A technique for determining laser profiles that produce
the required plasma excitation is developed.
We show that by properly shaping the longitudinal profile of the driving
laser pulse, it is possible to maximize both the transformer ratio and 
the wake amplitude, achieving optimal laser-plasma coupling.
The corresponding family of laser pulse shapes is derived in the 
nonlinear regime of laser-plasma interaction. 
Such shapes provide
theoretical upper limit on the magnitude of the wakefield and efficiency of
the accelerating stage by allowing for uniform photon deceleration inside the 
laser pulse. We also construct realistic optimal pulse shapes that 
can be produced in finite-bandwidth laser systems and propose a two-pulse 
wake amplification scheme using the optimal solution. 

\end{abstract}

\section*{Introduction}
Recent advances in laser technology allow one to create 
laser pulses with virtually arbitrary temporal intensity 
profiles using amplitude and phase shapers~\cite{shapers,murnane,downer}. 
Such laser pulses with  
non-Gaussian axial intensity are now being considered 
for applications as drivers in Laser Wakefield Accelerators (LWFA). 
Shaped lasers provide the means of controlling the generation of plasma 
wake and thus offer the possibility of optimization of wake excitation and 
accelerating efficiency. 
However, progress in finding ``the optimal'' shape has been hindered by the 
apparent complexity of the problem. Not only is the parameter space 
of possible shape functions huge, but also the generated wakefield is a nonlinear function 
of laser intensity, requiring numerical solution of differential equations 
in a variational calculation. As a result, 
several groups turned to trial and error methods such as genetic algorithms
for optimization \cite{murnane,downer}. 
Still, even these methods require consistent classification of laser shapes
so that different pulses can be meaningfully cross-compared while
 desired properties such as wake amplitude or efficiency are optimized.
In this paper we reanalyze the process of wake generation and
argue that the only two physical parameters that describe
a laser shape from the stand point of wake excitation are the total pulse 
energy and its depletion length. 
Using these parameters we find the {\it analytic} expression for the 
family of optimal laser shapes that maximize both the wakefield {\it and} 
the accelerating efficiency. We also develop a method for determining 
the shape of a laser that produces a required value of wakefield 
{\it without} explicitly solving the wake 
equation. This opens the way for obtaining laser shapes that satisfy other 
optimization criteria specific to given experimental conditions. 

\section*{Energy transfer in LWFA}

Wakefield accelerators such as the laser-driven LWFA~\cite{tajima} or electron beam driven Plasma
Wakefield Accelerator (PWFA) \cite{chen1} can be viewed as two-step 
energy transfer systems: 
in the first step the driver deposits energy into wake excitation of the plasma, 
and in the second step the energy is taken from the wake by the accelerating beam.
While the second step is the same for both accelerating schemes, the physics
of driver energy deposition is quite different between them.  
In PWFA the electron beam loses energy to the plasma through
 interaction with the induced electrostatic field, while in the 
LWFA laser energy loss occurs via
photon red-shift or deceleration \cite{wilks}. This process can be understood 
as follows. Poderomotive force of the laser modifies both the density $n_e$ and
the Lorentz factor $\gamma$ of the plasma electrons. This produces modulations in the
nonlinear index of the 
refraction $\eta\equiv [1-(\omega_p/\omega)^2 n_e/\gamma n_p]^{1/2}$,
where $\omega_{p} \equiv \sqrt{4 \pi e^2 n_p/m_e}$ is the unperturbed plasma frequency, and 
$\omega$ is the frequency of the laser. The wake-induced modulations of
the refractive index appear stationary in the reference frame comoving with the laser
and cause laser photons to red- or blue-shift depending on the sign 
of refractive index 
gradient \cite{mori1}. Due to negligible scattering in the setting of 
laser accelerators,
the photon number in the laser is essentially constant, and the 
energy deposition into the plasma
is therefore determined by the photon deceleration. To address this 
quantitatively we consider
a laser propagating along the $z$ axis with initial frequency $\omega_0 \gg \omega_{p}.$
In the laser comoving frame, the
plasma response can be written in
terms of the independent dimensionless variables
$\zeta=k_{p}(z-v_g t)$ and $\tau=k_{p} c t$, where
$k_{p}$ 
is the plasma wavenumber, and  $v_g \approx -c$ is the laser group
velocity (for convenience, the laser is moving in the negative $z$ 
direction). 
Introducing dimensionless normalized scalar and
vector potentials $\phi(\zeta)$ and $a(\zeta)$,
the parallel and perpendicular electric fields are
$E_\parallel =  -(m c^2 k_{p }/e) \partial \phi/
\partial \zeta$ and
${E_\perp}=-(mc/e) \partial a/\partial t=-(mc^2 k_{p
}/e) \partial a/\partial \zeta$. 
The wakefield generation equation can then be written as
\cite{sprangle2,esarey}:
\begin{equation}
{d^2 x \over d \zeta^2}= {n_e \over n_p}-1
={1\over 2}\Big({1+a^2(\zeta)\over x^2}-1\Big), \label{pot}
\end{equation}

\noindent where 
 $x\equiv 1+\phi$ is the modified electrostatic
potential, and $a^2(\zeta)$ is the dimensionless laser intensity 
envelope averaged over fast oscillations. Prior to the arrival of the laser 
the normalized wakefield 
${\cal{E}}\equiv e E_\parallel/m c \omega_{p}= -dx/d\zeta$ is
zero.
A formal solution for the electric field outside the laser
can be written as the first integral of  (\ref{pot}): 
$[{\cal{E}}^{out}(\zeta)]^2=-{(x-1)^2/ x}+
\int_{-\infty}^{\infty} {a^2 x'/{x}^2} d\zeta$, which reaches a maximum value at $x=1$:
\begin{equation}
[{\cal{E}}^{out}_{max}]^2=-
\int_{-\infty}^{\infty} {a^2(\zeta) \Big ({\partial \over \partial \zeta} 
{1\over x}} \Big ) d\zeta.
\label{eoutmax}
\end{equation}
This expression can be understood in terms of the deposition of laser 
energy into plasma. For this we use the formula for 
local frequency shift of laser photons obtained from the analysis of 
laser evolution equation \cite{frshift,AAC98}:

\begin{equation}
{\partial \omega \over \partial z}= -{1\over
2}{\omega_{p}^2 \over \omega}  k_{p} 
{\partial \over \partial \zeta} {n_e \over \gamma n_p} 
=-{\omega^2_{p}\over 2 \omega} 
k_{p}\Big({\partial \over \partial \zeta} {1\over x}\Big).
\label{enloss2}
\end{equation}
The energy density in the wake from (\ref{eoutmax}) can then be interpreted 
as the intensity-weighted integral of the photon deceleration throughout 
the pulse. Let's denote the wake-dependent part of the photon deceleration 
function as $\kappa(\zeta)\equiv x'/x^2$. The value of the peak wakefield 
in (\ref{eoutmax}) is then bounded from above 
by the  total laser energy (the integral of $a^2$) and the maximum photon 
deceleration $\kappa_{max}$:
\begin{equation}
[{\cal{E}}^{out}_{max}]^2=
\int_{-\infty}^{\infty} {a^2(\zeta) \kappa(\zeta)} d\zeta \leq
\kappa_{max} \int_{-\infty}^{\infty} {a^2(\zeta)} d\zeta ,
\label{wake2}
\end{equation}
 where $\kappa_{max}$ is the maximum of $\kappa(\zeta)$ 
inside the laser. 
Maximum photon deceleration $\kappa_{max}$ actually has a simple physical 
interpretation. It is closely related to the 
characteristic laser depletion length $l_d$, or the distance in which 
the maximally decelerated laser slice red-shifts down to $\omega_p$ 
(assuming no evolution of the wakefield). From (\ref{enloss2}) this
characteristic depletion length is:
\begin{equation}
l_d={[({\omega_0/ \omega_{p}})^2-1] / k_{p} \kappa_{max}} .
\end{equation}
The peak wakefield outside the laser then scales with depletion length $l_d$
and dimensionless pulse energy $\varepsilon_0\equiv \int_{-\infty}^{\infty} {a^2(\zeta)} d\zeta$
as: ${\cal{E}}^{out}_{max} \leq \sqrt{ [({\omega_0/ \omega_{p}})^2-1] \varepsilon_0/ k_p l_d}.$
The range of achievable wakefields is therefore set by the total pulse energy and its depletion
length. For pulses of fixed energy and depletion length the actual value of the wakefield
within this range will depend only on particular laser shape, and can be optimized by varying the 
shape subject to constraints.

\begin{figure}
\begin{center}
\epsffile{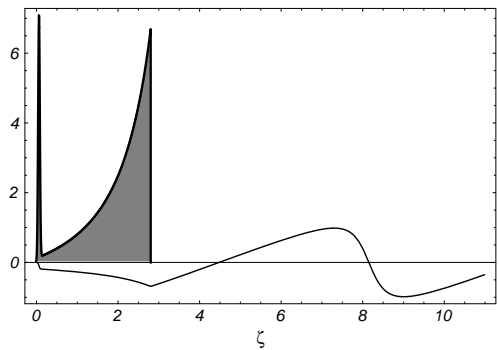}
\end{center}

\caption{General shape of the nonlinear optimal laser intensity
profile and its corresponding wakefield (arbitrary units)} \label{Shapefig}
\end{figure}

\section*{Wakefield optimization}

One possible optimization problem can be formulated as follows: 
given some fixed
laser energy $\varepsilon_0$
and the desired depletion length
($\kappa_{max}=\kappa_0$), what laser shape would produce the largest
possible wakefield? According to (\ref{wake2}) this amounts to finding a shape
that maintains the largest $\kappa(\zeta)$ for the duration of the pulse. 
Since the maximum $\kappa(\zeta)$ is fixed by the depletion length, such a shape
should have a constant photon deceleration throughout the pulse, $\kappa(\zeta)=\kappa_0$.
If the laser is present for $\zeta>0$, then in 
order to satisfy the boundary condition of quiescent plasma before the laser, 
the photon deceleration should  rise from $0$ to value $\kappa_0$ at the 
very beginning of the pulse, e.g.,
like a step-function: $\kappa(\zeta)=\kappa_0 \theta(\zeta^+)$. Here,
$\zeta^+\equiv \zeta-0^+$ in order to avoid ambiguities with
the values of step-function at 0. The corresponding laser profile is
then found from the wake equation (\ref{pot}):
\begin{equation}
a_l^2({\zeta})={2 {\kappa}_0 \delta({\zeta}^+) \over
(1-{\kappa}_0 {\zeta})^4}+{4{\kappa}_0^2 \theta^2({\zeta}^+)\over
(1-{\kappa}_0 {\zeta})^5}+{1\over (1-{\kappa}_0
{\zeta})^2}-1, \label{shape}
\end{equation}
where $ \zeta \in [0,\zeta_f<1/\kappa_0]$,
 and $\delta(\zeta^+)$ is a delta-function such that 
$\int_0^{\zeta>0}\delta(y^+)dy=1$.
A schematic drawing of the optimal laser intensity variation and
its associated plasma wakefield are shown in Fig.\ {\ref{Shapefig}}.
Generally, the shape consists of a $\delta$-function at the 
front followed by a ramp in intensity which is cut off at $\zeta_f$. 
In the linear regime, when $a^2 \ll 1$,
$\kappa_0\to 0$, the ramp reduces to a triangular shape 
found in \cite{AAC98,kyoto}:
$a^2={2\kappa_0}(\delta({\zeta^+})+{\zeta})$. 
We note that (\ref{shape}) describes a family of shapes, rather than a
fixed shape. 
The actual profile of the optimal pulse depends on the deceleration 
parameter $\kappa_0$ set by the desired depletion length and the pulse
 length $\zeta_f$, which is determined from the available total energy:
\begin{equation}
\varepsilon_0 = 
2 \kappa_0 + {\zeta_f [\kappa_0^2+(1-\kappa_0 \zeta_f)^3] \over
(1-\kappa_0 \zeta_f)^4}.
\label{energy}
\end{equation}
Although the pulse length cannot exceed $\zeta_c \equiv 1/\kappa_0$,
the rise of $a^2$ towards the end of the pulse
guarantees that any finite laser energy can be accommodated for
$\zeta_f < \zeta_c$.
The two terms in (\ref{energy}) represent the energy contained in the 
$\delta$-function precursor and the main pulse. It is clear that for a fixed
total energy there exists a maximum value of $\kappa_0=\varepsilon_0/2$ 
which is achieved
when $\zeta_f \to 0$, i.e., all of the energy is concentrated in the 
$\delta$-function. This shape, which is a particular case of the 
general optimal shape (\ref{shape}), excites the largest possible wakefield
and has the smallest depletion length among all pulses of fixed energy. 
For circularly polarized pulses with cylindrical transverse 
crossection of radius $r_0$ and wavelength $\lambda$, the maximum 
achievable wake is then given by:
\begin{equation}
E_{\mathrm{max}}=6.54 E_{wb} \Big[{U_0\over 1\mathrm{J}} \Big]\Big[{\lambda\over 1\mu 
\mathrm{m}} \Big]^2
\Big[{10 \mu \mathrm{m}\over r_0} \Big]^2 
\Big[{n_p \over 10^{18}\mathrm{cm}^{-3} } \Big]^{1/2}
\label{maxe}
\end{equation}
where $U_0$ is the total pulse energy (in Joules) and 
$E_{wb}=96 [n_p/10^{18} \mathrm{cm}^{-3}] \mathrm{GV/m}$ is the 
nonrelativistic wavebreaking field.

\section*{Efficiency optimization}
While generation of large accelerating gradients is a prerequisite for a 
successful accelerating scheme, the efficiency of acceleration should also
be considered. For an accelerating scheme that involves transfer of
 energy from 
the driver beam to the accelerating beam, efficiency is measured in terms
of the {\it transformer ratio}, or the ratio of the maximum rate of energy 
gain per particle of accelerating beam to the maximum rate
of energy loss per particle of the driving beam. In the case of laser-plasma
accelerators, where the driving and accelerating beams consist of particles 
of different species, the following kinematic definition is more useful:
\begin{equation}
R\equiv{|{\partial \gamma_a / \partial z}|_{{max}} 
\over |{\partial \gamma_d / \partial z}|_{{max}}}, \label{R1}
\end{equation}
where $\gamma_d$ and $\gamma_a$ are Lorentz factors for the driving and 
accelerating beams. In LWFA the particles in the trailing
electron bunch are accelerated via electrostatic interaction
with the wake, so ${|{\partial \gamma_a /
\partial z}|_{{max}}}  =
 |e E_\parallel^{{max}}|/m_e c^2=
 k_{p} |{\cal{E}}^{out}_{{max}}| $. For the laser propagating in plasma
$\gamma_d \approx \omega/\omega_p$, so  ${|{\partial \gamma_d /
\partial z}|}$ is the photon frequency shift given by (\ref{enloss2}). 
The transformer ratio for LWFA is then:
\begin{equation}
R^{{\mathrm LWFA}}=
{2 \omega\over \omega_{p}} 
{|{\partial x /\partial \zeta} |_{max}^{out} \over 
|{\partial ({1/ x})/ \partial \zeta} |_{max}^{in}}
 \propto {|\cal{E}|}^{out}_{max} k_{p} l_d .
 \label{R3}
\end{equation}
Defined this way  the transformer ratio can have several interpretations.
On the one hand, it is a measure of local accelerating efficiency, or 
the amount of increase in $\gamma$ of the accelerating electron per unit
loss of $\gamma$ of the laser. On the other hand, transformer ratio
is proportional to the maximum energy that can be transferred to  
the accelerating beam particle over the laser depletion length (assuming
no evolution of the wake during propagation). Therefore, an efficient 
accelerating scheme should attempt to maximize the transformer ratio.

There are several ways to find the laser shape that maximizes $R$. 
Among the pulses of fixed energy and depletion length, $R$ is maximized 
by a pulse that produces the largest wakefield as can be seen from (\ref{R3}). The 
optimal shape found in Eq. (\ref{shape}) satisfies this requirement. 
Alternatively, one can relax the energy constraint and instead look  
for a laser profile that has the largest depletion length among all the shapes 
that produce a given maximum wakefield behind the laser. Although
this reasoning leads to the same resulting shape, we include the proof for 
completeness as it demonstrates a useful technique for determining laser shapes
that satisfy constraints on the values of the wakefield.

In order to find the shape that maximizes the transformer ratio,  
 we vary the photon deceleration function $\kappa(\zeta)$ inside the laser.
We require that $\kappa(\zeta)$ be positive definite, i.e.,  
laser photons only {\it lose} energy to the plasma and do not reabsorb
energy from the wake. The advantage of varying
$\kappa(\zeta)$ rather than $a^2(\zeta)$ directly is that one can
immediately write down the solution for the wakefield potential
$x(\zeta)$ in terms of the photon energy deposition function
$\psi(\zeta) \equiv \int_{-\infty}^\zeta \kappa(\zeta_1) d \zeta_1$, i.e.,
$x(\zeta)=1/(1-\psi(\zeta))$. The corresponding laser shape is then 
determined from the wakefield equation (\ref{pot}):

\begin{eqnarray}
a^2(\zeta)= (2 x''(\zeta)+1) x(\zeta)^2-1=
{2 \psi''(\zeta) \over [1-\psi(\zeta)]^4}
+{4 \psi'(\zeta)^2 \over [1-\psi(\zeta)]^5} +{1\over[1-\psi(\zeta)]^2}
-1.
\label{las1}
\end{eqnarray}

Note that not all functions $\psi(\zeta)$
should produce physical, i.e., positive $a^2(\zeta)$, and the validity
of a given $\kappa(\zeta)$ should be checked through (\ref{las1}).
By considering photon energy deposition in the pulse 
all possible laser shapes 
that produce a given wakefield 
can be mapped onto a bounded space of monotonically increasing
functions $\psi(\zeta)$, whose end values on the interval
$[0,\zeta_f]$ and derivatives at $\zeta_f$ are constrained by the required
maximum value of the wakefield. From the first integral of the wakefield
equation we can relate the modified potential $x_f\equiv
x(\zeta_f)$
 and the electric field $x'_f\equiv x'(\zeta_f)$ 
evaluated at the end of the pulse:

\begin{equation}
({\cal{E}}^{out}_{max})^2={(x_f-1)^2\over x_f}+(x'_f)^2 
={\psi(\zeta_f)^2 \over (1-\psi(\zeta_f))}+{\psi'(\zeta_f)^2\over
(1-\psi(\zeta_f))^4}. \label{constr}
\end{equation}

\begin{figure}
\begin{center}
\epsfxsize=5.845in
\epsfysize=1.75in
\epsffile{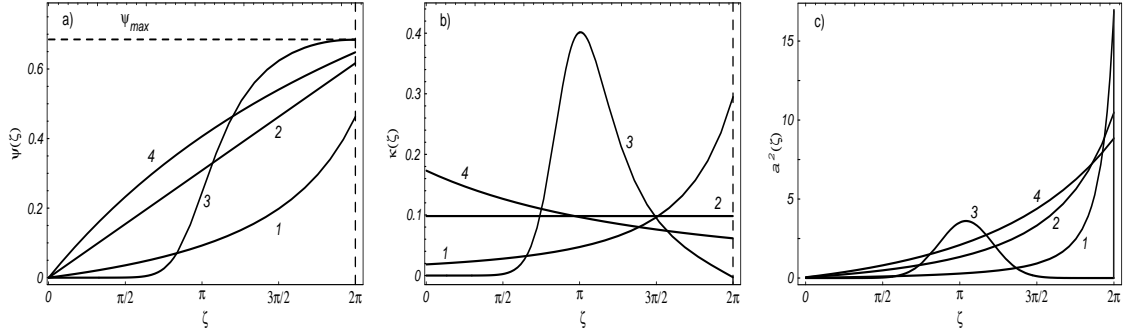}
\end{center}
\caption{a) Sample photon energy deposition 
$\psi(\zeta)$ for pulses of length not exceeding $\zeta_f=2 \pi$; b)
corresponding photon deceleration functions; c) resulting laser
intensity profiles. All shapes produce the same maximum wakefield. }
\label{psifig}
\end{figure}

Monotonicity of $\psi(\zeta)$ follows from the requirement 
$\kappa(\zeta)\ge 0$,
and the bounds on $\psi(\zeta)$ are $0=\psi(0)\leq \psi(\zeta) 
\leq \psi_{max}<1$. A few sample solutions for $\psi(\zeta)$ and 
corresponding photon deceleration and laser shapes are plotted 
in Fig.\ \ref{psifig}.
The function $\psi(\zeta)$ that results in the largest transformer ratio
should possess the smallest maximal slope in the interval $[0,\zeta_f]$ --
this will maximize the depletion length for a fixed ${\cal{E}}^{out}_{max}$.
Such curve is unique and is represented by curve 2 in figure
\ref{psifig}. It is a straight line with slope
$\psi'(\zeta_f)={\psi(\zeta_f)_{\mathrm{crit}}/ \zeta_f}$, where
the value of $\psi(\zeta_f)_{\mathrm{crit}}$ is determined from
substituting $\psi'(\zeta_f)$ into eq. (\ref{constr}). Let's
show that this line has the smallest maximum slope. Since
$\psi'(\zeta_f)$ is a decreasing function of $\psi(\zeta_f)$ for 
a fixed ${\cal{E}}^{out}_{max}$ (eq. (\ref{constr})), all curves
$\psi(\zeta)$ with $\psi(\zeta_f)< \psi(\zeta_f)_{\mathrm{crit}}$
(such as curve 1 in figure \ref{psifig}) will automatically have
larger slope at $\zeta_f$:
$\psi'(\zeta)>\psi(\zeta_f)_{\mathrm{crit}}/\zeta_f$. On the other hand, 
the curves
with $\psi(\zeta_f)> \psi(\zeta_f)_{\mathrm{crit}}$ (such as
curves 3 and 4) should have a slope larger than
$\psi(\zeta_f)_{\mathrm{crit}}$ somewhere between $0$ and
$\zeta_f$ in order to be larger than $\psi(\zeta_f)_{\mathrm{crit}}$ at
$\zeta_f$. We therefore prove that the
 function $\psi(\zeta)=\kappa_0 \zeta$, where
$\kappa_0 \equiv \psi(\zeta_f)_{\mathrm{crit}}/\zeta_f$, is an
integrated photon deceleration profile that maximizes the
transformer ratio. 

The photon deceleration function associated with this $\psi(\zeta)$
is a constant $\kappa(\zeta)=\kappa_0$ and the resulting laser shape
is the same as given by (\ref{shape}). The optimal transformer ratio
associated with this shape can be found from (\ref{R3}):
\begin {equation}
R^{{\mathrm LWFA}}={2\omega\over \omega_{p}}
\sqrt{{1+(k_{p} L_p)^2 [1-{\kappa_0} (k_{p} L_p)]^3
\over [1- {\kappa_0} (k_{p} L_p)]^4}},
\label{TRopt} 
\end{equation}
where $L_p=\zeta_f/k_p$ is the pulse length.
In the linear regime optimal transformer ratios for both
LWFA and PWFA schemes scale identically with the pulse/beam length:
$R^{{\mathrm LWFA}}\to ({2 \omega / \omega_{p}})
\sqrt{1+(k_{p} L_{pulse})^2}, $ $R^{{\mathrm PWFA}} \to 
\sqrt{1+(k_{p} L_{beam})^2}$ \cite{chen2}. 
The LWFA scheme is intrinsically more efficient
by a factor of $2 \omega/\omega_{p}$, which is needed for viability of LWFA
 since lasers are
typically ``slower'' drivers than electron beams.

\section*{Utility of pulse shaping}

\begin{figure}
\unitlength = 0.0011\textwidth
\begin{center}
\epsffile{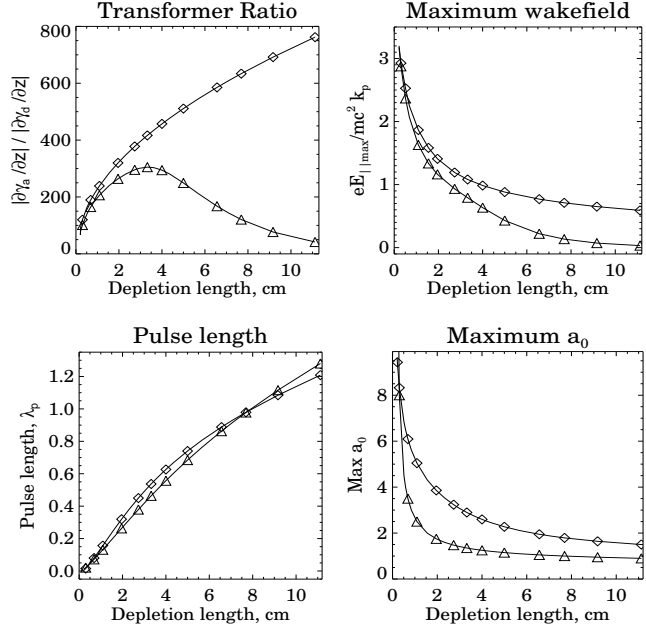}
\end{center}
\caption{
Comparison of the transformer ratio, maximum wakefield, pulse length, and
maximum normalized vector potential in shaped (diamonds) and Gaussian (triangles)
pulses of equal depletion lengths and constant pulse energy of $0.5 {\mathrm{J}}$.}
\label{TRcompar}
\end{figure}

The advantage of using the optimal pulse shape is best seen in
comparison with the unshaped (Gaussian) pulse. 
For a given
Gaussian pulse (or any other non-optimal shape) one can always
construct a corresponding optimally shaped pulse with the same laser 
energy such that the
photon deceleration across the optimal pulse equals to the peak photon
deceleration in the unshaped one 
(i.e., both pulses have equal depletion lengths). 
Unshaped pulses deplete first in
the region where photon deceleration is the largest, whereas a laser
with the optimal shape 
loses {\it all} its energy in a depletion length due to uniform 
photon deceleration, 
thus enhancing instantaneous energy deposition and wakefield.
For a
numerical example, we consider the optimal and Gaussian pulses of
total energy $0.5\mathrm{J}$, wavelength $1 \mu \mathrm{m}$ and cylindrical
radius $10 \mu \mathrm{m}$ in a plasma with $n_p=10^{18}
{\mathrm{cm}}^{-3}$.
 The transformer ratio, the maximum wakefield,
the required pulse length, and the corresponding peak $a_0$ are shown in 
Fig.~\ref{TRcompar} as a function of depletion length.

From Fig.\ \ref{TRcompar} we see that the transformer ratio and the
maximum wakefield are consistently larger for shaped
 pulses. In fact, the lines for optimal pulse wakefield and transformer ratio
 represent 
the theoretical upper limits for all pulses of given energy.
The Gaussian pulse achieves a maximum transformer ratio when its
length (measured here as FWHM) equals $1/2$ of the relativistic plasma wavelength.
The effects
of shaping are especially prominent for longer pulses, where
Gaussian pulse yields almost no wake excitation due to plasma
oscillations inside the pulse that cause part of the laser photons
to absorb energy from the wake. On the other hand, a shaped
laser postpones plasma oscillation until the end of the pulse, and 
all photons decelerate uniformly.
 For very short pulses, the differences between the 
two shapes are minimal. This
is due to the fact that very short Gaussian pulses of fixed energy 
asymptotically approach the delta-function limit of the short optimal 
shape. For these short pulses the wakefield attains the maximum value
given by (\ref{maxe}) as the depletion length reaches the minimal value 
for given pulse energy.

Although short pulses generally produce the largest wakefields, their efficiency
is close to minimal possible, as the depletion length decreases
faster than increase in the wake. Therefore, the choice of the 
appropriate pulse shape for LWFA stage will depend on specific
experimental conditions. If the laser-plasma interaction distance is limited
by instabilities, diffraction or dephasing, then in order to maximize 
the electron
energy gain one should try to achieve the largest accelerating gradient, which
can be accomplished with ultrashort pulses. 
For some regimes of plasma density and laser energy available laser systems
may be unable to produce pulses short enough so that the pump depletion length 
is longer than the characteristic instability or dephasing length. In this case
shaping the laser will increase the wakefield over the
interaction distance, even though it will be below the maximum possible if
a shorter pulse were used. 
If the interaction length is less constrained, such as the case for 
propagation in plasma channels \cite{channels}, 
then using a finite-length shaped pulse
 will result in a greatly improved overall energy gain per stage as can be seen from Fig.\ \ref{TRcompar}.
An added benefit of pulse shaping is the suppression of modulational
instability that affects unshaped pulses that are longer than 
plasma wavelength. When
all photons red-shift, or ``slow down'', at the same rate,
different slices of the laser do not overrun each other, and the
1D laser self-modulation is suppressed.

\section*{Realistic pulse shaping}
As the optimal pulse shape is
associated with a delta-function precursor, the feasibility of such
a structure may be a concern. We note that the purpose of this
precursor is to bring the photon
deceleration from zero in the quiescent plasma before the laser
to a finite value $\kappa_0$ at the beginning of the main pulse. This
 can also be achieved with a more physical prepulse, whose shape
can be found from the wake equation once a smooth function $\kappa(\zeta)$ is
chosen. 

\begin{figure}
\unitlength = 0.0011\textwidth
\begin{center}
\epsffile{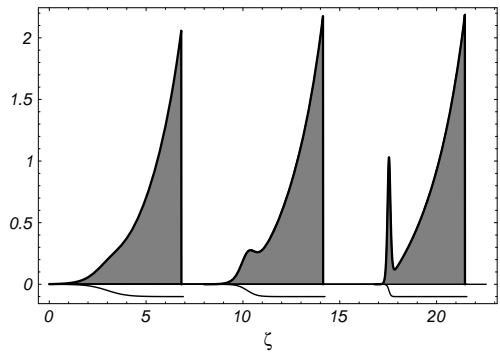}
\end{center}
\caption{Laser intensity (shaded) and associated photon deceleration 
($-\kappa(\zeta)$)
for pulses of the same total energy and characteristic depletion length 
in the order of increasing $\alpha$. }
\label{multishape}
\end{figure}

For our example we choose a photon deceleration function that varies
as a hyperbolic tangent: $\kappa(\zeta)=\kappa_0 [1+\tanh (\alpha (\zeta-\zeta_0))]/2$,
where $\alpha$ is a steepness parameter and $\zeta_0$ is an arbitrary offset. 
The photon energy deposition is then 
$\psi(\zeta)=\kappa_0 [\zeta+\ln(\cosh(\alpha (\zeta-\zeta_0)))/\alpha]/2$, and the corresponding
laser shape is found from equation (\ref{las1}):
\begin{equation}
a^2(\zeta)={\kappa_0 \alpha {\mathrm sech}^2(\alpha (\zeta-\zeta_0)) \over \chi^4(\zeta)}+
{\kappa_0^2 [1+\tanh(\alpha (\zeta-\zeta_0))]^2 \over \chi^5(\zeta)}+{1\over \chi^2(\zeta)}-1,
\label{newlas}
\end{equation}
where $\zeta \leq \zeta_f$ and the function in the denominator is 
$\chi(\zeta)=1+(\kappa_0 / 2 \alpha)\ln{1\over 2}+\zeta_0 \kappa_0/ 2 -\psi(\zeta)$.
As before, the pulse length $\zeta_f$ can be found from the total available energy of the pulse.  
By varying $\alpha$ we can change the slope of $\kappa(\zeta)$ as it rises from $0$ to $\kappa_0$
and construct a pulse shape that satisfies experimental constraints yet retains essential 
physics of the optimal shape. 
For a step-function photon deceleration ($\alpha \to \infty$) expression (\ref{newlas}) asymptotes
to equation (\ref{shape}). However, for finite values of $\alpha$ the delta-function precursor spreads out
and can even disappear as shown in Fig.~{\ref{multishape}}. 
The family of shapes given by (\ref{newlas}) is better suited for the finite-bandwidth laser systems that 
have a lower limit on achievable feature size. The values of maximum wakefield 
for pulses in Fig.~{\ref{multishape}} is within few percent of the value for a delta-function
optimal pulse of the same energy and depletion length. 
This is due to the fact that the bulk of the laser pulse still experiences constant maximal
photon deceleration. The wakefield further degrades with longer rise times of $\kappa(\zeta)$.

\begin{figure}

\begin{center}
\epsffile{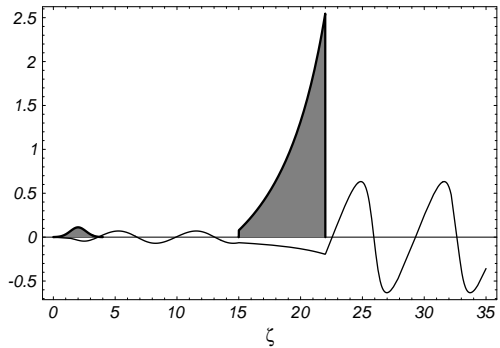}
\end{center}

\caption{Laser intensity profiles ($a^2(\zeta)$, shaded) and normalized electric field
for optimally shaped main pulse following a Gaussian precursor.
 }
\label{twopulse}
\end{figure}

The pulse shaping techniques described so far have assumed that the laser is
incident on an unperturbed plasma. However, this does not have to be the case,
and we can construct an optimally-shaped laser that enters the plasma at some
phase of a pre-existing plasma oscillation.  Such oscillation could be left
from a precursor laser pulse or electron beam as shown in Fig.~\ref{twopulse}.
When there is an existing plasma wave, the value $x_0$ of the modified
electrostatic potential at the beginning of the optimal pulse will generally be
different from unity. In this case the expression for the optimal pulse without
the delta-function precursor is modified into:

\begin{equation}
a_l^2 ({\zeta})
={4{\kappa_0}^2 \over
[{ x_0^{-1}}-{\kappa_0} {(\zeta-\zeta_0)}]^5}+{1\over [{ x_0^{-1}}-{\kappa_0}
 {(\zeta-\zeta_0)}]^2}-1 , 
\label{shape1}
\end{equation}
where we assume that the main pulse lies between $\zeta_0$ and $\zeta_f$,
and $\kappa_0=x'(\zeta_0)/x_0^2$.
If this shape is placed in a correct phase of the oscillation (so that 
$a^2_l(\zeta_0)$ from (\ref{shape1}) is positive), it
acts as an amplifier of the existing wakefield. 
The ratio of maximum wakefield 
behind the optimal pulse to the field
in front of it scales as $(R/x_0^2)(\omega_p /2 \omega)$ which for pulse 
lengths around $\lambda_p$ from Fig.~\ref{TRcompar} can be of order $10$. 
A detailed discussion of
this scheme and a comparison to the resonant laser-plasma
accelerator concept~\cite{RLPA} will be reported elsewhere.

\section*{Discussion}
As we have shown, the huge phase space involved in shaping laser
drivers for applications in laser wakefield accelerators can be
described using only two parameters: total pulse energy and characteristic
depletion length. The shape of photon energy deposition (photon
deceleration) inside the pulse plays a crucial role for both the wake
excitation and the evolution of the laser driver. By varying the shape of
the photon deceleration function for pulses of fixed energy and depletion 
length we were able to optimize both the generated wakefield and the
efficiency of the accelerating scheme.
The method used for obtaining the optimal shapes (\ref{shape})
and (\ref{newlas})
 is actually more general and can be used to determine laser shapes 
that 
 generate other variations in the nonlinear index of refraction.
Having a physical requirement for the refractive index, 
which in this case is the requirement of uniformity
of photon deceleration, provides a constraint on the functional
form of the wakefield, which can then be
used to find the required laser shape.
Alas, such a ``reverse'' solution is not always guaranteed to yield a 
physical 
(i.e., positive) $a^2(\zeta)$, so, in general, caution is advised.

Several issues should be addressed before
the laser pulse shaping concept can be fully utilized.
Even without the delta-function precursor, the finite laser bandwidth 
will necessarily smooth out steep rises and falls of the 
optimal pulse shape. Although we do not anticipate adverse effects when
the feature size is much smaller than the plasma wavelength, the 
1D self-consistent laser evolution and stability of realistic 
optimal shapes are currently under investigation. 
 Another
consideration is the influence of the laser-plasma interaction in
the transverse dimension on the evolution of the pulse. 
Many of the laser-plasma instabilities are seeded by the wakefield-induced
perturbations of the index of refraction. As we have demonstrated in this 
paper, the nonlinear index of refraction can be effectively 
controlled through
laser shaping, thus suggesting the method of delaying the onset of 
these instabilities. Whether this approach increases the growth rates 
of other instabilities, particularly in the transverse dimension, remains to
be investigated.

We would like to thank J. Arons, A. Charman, T. Katsouleas, W. B. Mori, and 
J. Wurtele for fruitful discussions and suggestions.


\end{document}